\documentclass{elsarticle}

\usepackage[left=3cm,right=3cm,top=3cm, bottom=4cm]{geometry}
\usepackage{amsmath,amssymb,amsfonts}
\usepackage{algorithmic}
\usepackage[ruled,vlined, linesnumbered]{algorithm2e}
\usepackage{graphicx}
\usepackage{textcomp}
\usepackage{bm}
\usepackage{xcolor}

\def\BibTeX{{\rm B\kern-.05em{\sc i\kern-.025em b}\kern-.08em
    T\kern-.1667em\lower.7ex\hbox{E}\kern-.125emX}}

\usepackage{hyperref}
\usepackage{verbatim} 

\usepackage{graphicx}
\usepackage{caption}
\usepackage{subcaption}
\usepackage{amssymb}
\usepackage{ textcomp } 
\usepackage{color}
\usepackage{enumerate}
\usepackage{mathrsfs}
\usepackage{multicol}
\usepackage{hyperref}
\usepackage{amsopn}

\usepackage{amsmath}
\usepackage{amsthm}
\usepackage{amssymb}

\usepackage{multicol}

\newcommand{\C}{\mathbb{C}}

\newcommand{\K}{\mathbb{K}}
\newcommand{\N}{\mathbb{N}}

\newcommand{\R}{\mathbb{R}}

\newcommand{\abs}[1]{\left\vert #1 \right\vert}
\newcommand{\norm}[1]{\Vert #1 \Vert}

\newcommand{\set}[1]{\left\lbrace #1\right\rbrace}
\newcommand{\sse}{\subseteq}
\newcommand{\sprod}[1]{\left\langle #1 \right\rangle}

\newcommand{\prb}[1]{\mathbb{P}\left( #1 \right)}

\usepackage{dsfont}

\newcommand{\geqsim}{\gtrsim}
\newcommand{\leqsim}{\lesssim}

\DeclareMathOperator{\supp}{supp}

\newcommand{\argmin}{\mathop{\mathrm{argmin}}}

\DeclareMathOperator{\tr}{tr}

\newtheorem{lem}{Lemma}
\newtheorem{prop}[lem]{Proposition}

\newtheorem{theo}[lem]{Theorem}

\newtheorem{defi}{Definition}
\newtheorem{rem}{Remark}

\theoremstyle{definition}

\numberwithin{lem}{section}

\usepackage{etoolbox}
\let\bbordermatrix\bordermatrix
\patchcmd{\bbordermatrix}{8.75}{4.75}{}{}
\patchcmd{\bbordermatrix}{\left(}{\left[}{}{}
\patchcmd{\bbordermatrix}{\right)}{\right]}{}{}

\renewcommand{\vec}[1]{{\bf #1}}

\newcommand{\af}[1]{{#1}}

\definecolor{ingo}{rgb}{1,.2,.4}

\definecolor{bblue}{rgb}{0,0,0.8}

\newcommand{\sigmas}{\bm{\sigma}}

\usepackage{xfrac} 

\title{Hierarchical Isometry Properties of Hierarchical Measurements \tnoteref{t2}}

\tnotetext[t2]{Part of this work has been submitted to the IEEE Statistical Signal Processing Workshop 2021.}

\author[2]{Axel Flinth\corref{cor1}}
\author[1]{Benedikt Gro\ss}
\author[3,4]{Ingo Roth}
\author[3]{Jens Eisert}
\author[1]{Gerhard Wunder}

\address[2]{Institute for Electrical Engineering, Chalmers Institute of Technology, Hörsalsvägen 11, 412 58 Gothenburg, Sweden}
\address[1]{Department of Computer Sciences, Freie Universität Berlin, Takustra\ss e 9, 14195
Berlin, Germany}
\address[3]{Institute for  Theoretical Physics, Freie Universität Berlin, Arnimallee 14, 14195 Berlin, Germany}
\address[4]{Quantum Research Centre, Technology Innovation Institute, Abu Dhabi, UAE}

\cortext[cor1]{Corresponding author.}

\begin{document}

\begin{abstract}
  Compressed sensing studies linear recovery problems under structure assumptions. 
  We introduce a new class of measurement operators, coined hierarchical measurement operators, and prove results guaranteeing the efficient, stable and robust recovery of hierarchically structured signals from such measurements.  
  We derive bounds on their hierarchical restricted isometry properties based on the restricted isometry constants of their constituent matrices, generalizing and extending prior work on Kronecker-product  measurements. As an exemplary application, we apply the theory to two communication scenarios.  The fast and scalable HiHTP algorithm is shown to be suitable for solving these types of problems and its performance is evaluated numerically in terms of sparse signal recovery and block detection capability.
\end{abstract}

\maketitle


\section{Introduction}

The general idea of \emph{compressed sensing}   \citep{candes2008introduction} is to exploit sparsity of a signal $\vec{x}$ to facilitate its recovery from incomplete and noisy linear measurements. 
The measurements being incomplete, the recovery problem is a priori ill-posed. 
Yet, if the signal is assumed to be sparse, efficient recovery is still possible. It can furthermore be theoretically guaranteed, provided the measurement operator $\vec{A}$ satisfies the \emph{restricted isometry property (RIP)} \cite{Can08}. We say that a matrix $\vec{A}$ has the $s$-RIP, if, for some constant $\delta_s(\vec{A})>0$,
\begin{equation}
\label{eq:standard_rip}
    |\|\vec{A\vec{x}}\|^2-\|\vec{x}\|^2| \leq \delta_s(\vec{A})\|\vec{x}\|^2
\end{equation}
for all $s$-sparse vectors $\vec{x}$  
\af{and with $\|\cdot\|$ denoting the (Euclidean)  $\ell_2$-norm}.
Sometimes, signals fulfil stronger structural assumption than sparsity. In this work, we consider \emph{hierarchical sparsity} (hi-sparsity) as such an assumption. We define hi-sparsity as follows \citep{FriedmanEtAl:2010,HiHTP,SimonEtAl:2013, SprechmannEtAl:2010,  SprechmannEtAl:2011}.

\begin{defi}[$(s, \sigmas)$-sparsity]
Let $\vec{x} = (\vec{x}_1, \dots, \vec{x}_N) \in \K^{n_1} \times \dots \times \K^{n_N}$, where $\K$ is either $\R$ or $\C$. For  $s$ and $\sigmas=(\sigma_1, \dots, \sigma_N)$, we say that $\vec{x}$ is $(s, \sigmas)$-sparse, if
\begin{itemize}
    \item at most $s$ blocks $\vec{x}_i$ are non-zero and
    \item each non-zero block $\vec{x}_i$ is $\sigma_i$-sparse.
\end{itemize}
\af{To simplify notation, we use the shorthand $(s,\sigma)$-sparsity for the case when $\sigma_i = \sigma$ for all $i$ and scalar $\sigma$.}

\end{defi}

\af{\begin{rem}The definition can be recursively generalized to more sparsity levels with a nested tree structure. To be concrete, instead of assuming that the non-vanishing blocks are simply $\sigma_i$-sparse, we can require each block $\vec x_i$ to consist of only $\sigma_i$ non-vanishing sparse blocks. In other words, we require $\vec x_i$ to be  
 $(\sigma_i, \bm{\varsigma}_i)$-sparse, with a vector  $\bm{\varsigma}_i=(\varsigma_{i,1}, \dots, \varsigma_{i,n_i})$ specifying the sparsity level of the blocks of the block $\vec x_i$.  
 This gives rise to a three-level hierarchically structured vector. 
 Recursively nesting the definition, we can organize the entries of a vector in a rooted tree of blocks with a sparsity restriction on each level, we refer to Ref.\ \citep{HiHTP} for details. 
 Importantly, most of our results hold for such general hierarchically sparse vectors with multiple levels of blocks. 
 The proofs can be verbatim translated by simply allowing each $\sigma_i$ to refer to an arbitrary multi-level hierarchical sparsity structure. 
 We will however refrain from highlighting this in our notation to keep it concise. \label{rem:multilevel}
\end{rem}}

Hi-sparse signals indeed appear in many applications. As a motivating example, we consider sporadic communication of a massive number of devices in the Internet of Things (IoT). In this scenario, we can imagine a large set of $N$ devices sending messages $\vec{x}_i$ to a base station. It is common to assume a sporadic device activity, i.e.,\ that only a few devices are active at each instant. This corresponds to only a few blocks being non-zero. Additionally assuming that the messages are sparse (or sparsely encoded), the vector $\vec{x}=(\vec{x}_1, \dots, \vec{x}_n)$ of all messages that the base station needs to recover becomes hierarchically sparse, see for an overview, e.g., Ref.\ \cite{Bockelm2018_ACCESS}. \af{ Alternatively, the messages themselves could also be assumed to be hierarchically sparse. For instance, wireless signals are often sparse in the \emph{angle-delay domain}, which can be modelled as a type of hierarchical sparsity (see Ref. \ \cite{Wunder_TWC19}). With such an assumption, $\vec{x}$ becomes hierarchically sparse in three levels (see Remark \ref{rem:multilevel}).}

 A subset of the authors of this work has introduced a general RIP-based recovery framework for hierarchically sparse vectors 
 in Refs.\  \citep{HiHTP,RothEtAl:iTwist:2016}, serving as our starting point here. In particular, they have introduced the \emph{hierarchical HTP (HiHTP)}, an adapted version of the celebrated 
\emph{hard threshold pursuit (HTP)} 
\citep{Foucart:2011} to recover hi-sparse signals. 
Importantly, for hierarchical sparsity, the projection step at the core of hard-thresholding algorithms, can be computed in the same computational complexity as sparse hard-thresholding. For this reason, HiHTP is efficient. 
The pseudo-code of the algorithm is given in Section~\ref{sec:numerics}. 
HiHTP comes with a recovery guarantee based on an adapted version of the RIP, the \emph{hierarchical RIP (HiRIP)}.
\begin{defi}[$(s,\vec{\sigma})$-HiRIP constant]
    Let $\vec{A}\in \K^{\tilde m,\tilde n}$. The smallest  $\delta > 0$ for which
\begin{align*}
    (1-\delta) \norm{\vec{x}}^2 \leq \norm{\vec{A}\vec{x}}^2 \leq (1+\delta) \norm{ \vec{x}}^2 
\end{align*}
for all $(s,\vec{\sigma)}$-sparse $\vec{x}$ is called the $(s,\vec{\sigma})$-HiRIP constant of $\vec{A}$, $\delta_{s,\vec{\sigma}}(\vec{A})$.
\end{defi}
Provided the measurement operator has the HiRIP for suitable parameters, HiHTP can recover any hierarchically sparse signal in a stable and robust fashion. 
\af{To be concrete, the following result holds.
\begin{theo}  \cite[Theorem 9,  simplified version]{HiHTP} Suppose that $\delta_{(3s,3 \sigmas)}(\vec{A}) \leq 1/\sqrt{3}$. Given an $(s,\sigmas)$-sparse $\vec{x}$,
the iterates $\vec{x}^{(t)}$ of the HiHTP (Algorithm 1) with input data $\vec y = \vec A\vec{x} + \vec {e}$ with additive noise $\vec e$ satisfies
$
    \norm{\vec{x}^k-\vec{x}} \leq \rho^k \norm{\vec{x}^0-\vec{x}} + \tau \norm{\vec{e}}
$ 
for constant $\rho < 1$  and $\tau$
only dependent on $\delta_{(3s,3\sigmas)}(\vec{A})$.
\end{theo}
}

In the following, we simply say that a matrix `has the HiRIP' when $\delta_{s,\sigmas}(\vec{A})$ is small enough for the required parameters $s$ and $\sigmas$ such that the recovery can be guaranteed.

These results motivate the study of the HiRIP properties of measurement operators. 
Within the framework of model-based compressed sensing \cite{BarCevDua10}, HiRIP properties of Gaussian matrices can be directly derived by counting the number of sub-spaces constituting the signal structure \cite{HiHTP}. 
Going significantly beyond these standard arguments and intimately related to the the hierarchical structure, 
one can also establish a suitable HiRIP for Kronecker-product operators \cite{roth2018hierarchical,HiHTP}. 
In this work, we follow-up and fully explore the relation between hierarchically structured signals and correspondingly structured measurement operators.

We study a general class of structured operators that 
are aligned with the hierarchical block-structure of the signals. 
Accordingly, we refer to them as  \emph{hierarchical measurement operators}.

\begin{defi}[Hierarchical measurement operator]
	We call a measurement operator $$\vec{H}: \bigoplus_{i=1}^N \K^{n_i} \to \K^{M} \otimes \K^m$$ a \emph{hierarchical measurement operator} if there exists matrices $\vec{B}_i \in \K^{m\times n_i}$ and a matrix $\vec{A} \in \K^{M\times N}$ with \af{columns $\vec{a}_i\in \K^M$, $i\in [N] := \{1, 2, \ldots, N\}$} such that 
 	\begin{align}
		\vec{H}\left(\vec{x}_1, \dots, \vec{x}_N\right) = \sum_{i=1}^N \vec{a}_i \otimes (\vec{B}_i \vec{x}_i)\,. \label{eq:hierarchical}
	\end{align}
\end{defi}
We will refer to $\vec{A}$ as the \emph{top-level matrix} and the $\vec{B}_i$ as the \emph{sub-level matrices}.
\begin{rem} 
1. Kronecker product operators are special cases of hierarchical measurement operators. Indeed, if all $\vec{B}_i$ are equal to a single, common $\vec{B}$, the corresponding hierarchical operator is \af{equal to the operator given by the Kronecker product $\vec{A} \otimes \vec{B}$;
\begin{align*}
    (\vec{A} \otimes \vec{B})\vec{x} &= \begin{bmatrix}
    a_{1,1} \vec{B} & \dots  & a_{1,N}\vec{B} \\
    \vdots & \ddots & \vdots \\
    a_{M,1} \vec{B} & \dots & a_{M,N} \vec{B}\end{bmatrix}\begin{bmatrix}
    \vec{x}_1  \\
    \vdots  \\
    \vec{x}_N\end{bmatrix}= \begin{bmatrix}
    a_{1,1} \vec{B}\vec{x}_1 + \dots  +  a_{1,N}\vec{B}\vec{x}_N \\
    \vdots  \\
    a_{M,1} \vec{B}\vec{x}_1 +\dots + a_{M,N} \vec{B}\vec{x}_N\end{bmatrix} \\
    &= \sum_{i=1}^N \begin{bmatrix}
    a_{1,i} \vec{B}\vec{x}_i \\
    \vdots  \\
    a_{M,i} \vec{B}\vec{x}_i \end{bmatrix} = \sum_{i=1}^N \vec{a}_i \otimes (\vec{B}\vec{x}_i)\, .
\end{align*}}
2. Note that the $\vec{B}_i$ may be of different sizes, as long as they map into the same space.
\end{rem} 
A hierarchical measurement operator can be thought of as a `multi-shot mixing operator'. 
\af{For each `shot' $j\in [M]$, the  linear combination $\sum_{i=1}^N a_{j,i}\vec{B}_i\vec{x}_i$, is a `mixture' of the vectors} $\vec{B}_i\vec{x}_i \in \K^m$. This makes the structure relevant for applications.
For a concrete example, let us return to the IoT scenario that was briefly discussed above. We assume that each user employs a linear encoding operator $\vec{B}_i \in \C^{m,n_i}$ to encode his/her sparse message $\vec{x}_i \in \C^{n_i}$ into a sequence $\vec{B}_i \vec{x}_i \in \C^{m}$, and that the base station has $M$ antennas at its disposal. Over the course of $m$ time-slots, the $j$:th antenna will then receive 
\begin{align}
    \vec{y}_j = \sum_{i=1}^N a_{j,i}\vec{B}_i\vec{x}_i \label{eq:sparseCoding}
\end{align}
where $a_{j,i}$ is the complex channel gain from the $i$-th user to the $j$-th antenna. Hence, the entirety of the base station's measurement will be a hierarchical measurement of the vector $\vec{x}$ to be recovered.

\paragraph{Contributions}
By the results explained above, the recovery of hi-sparse signals can be guaranteed if a measurement operator has the HiRIP. 
Hierarchical measurement operators consist out of constituent matrices that act in alignment with the block-structures.  
On each level, \af{each block of a hi-sparse signal is} assumed to be sparse. A matrix with the RIP acts almost isometrically on such \emph{blocks}.
This motivates the main question of this work: How are the HiRIP constants of hierarchical operators related to standard RIP constants and coherence measures of the constituent matrices?

We drive two main results: The first one, Theorem~\ref{th:HiRIP}, 
establishes that the hierarchical measurement operator inherits RIP properties from its constituent matrices. 
If the top-level matrix $\vec{A}$ has the $s$-RIP and all sub-level matrices $\vec{B_i}$ have the $\sigma_i$-RIP, the hierarchical measurement operator will have the $(s,\sigmas)$-RIP. 

As a second result, Theorem \ref{th:Incoherent Blocks},
we show that the RIP assumption on the top-level matrix can be relaxed if the sub-level matrices are mutually incoherent in a specific notion. This provides a more detailed picture of HiRIP arising from the properties of the constituent matrices and introduces considerable flexibility to derive HiRIP for specific instances of hierarchical measurement operators. 
In addition to the analytical results, we study the recovery performance of hierarchical sparse signals from hierarchical measurement operators in numerical simulations. 
The numerical results on the one hand illustrate and verify our theoretical \af{results}. On the other hand, they showcase the applicability of our framework in two specific applications from mobile communications.

\paragraph{Prior work}

The concept of hierarchical sparsity falls into the framework of \emph{model-based compressed sensing} \cite{BarCevDua10}. 
{
The sparsity model of model-based compressed sensing is very general: A union of sub-spaces $\bigcup_{i} U_i \sse \K^n$ \cite{LuDo:2008:ATheory}%
, is distinguished as the subset of structured signals. 
Many standard recovery algorithms can be generalized to this sparsity model. 
In order to adapt greedy approaches such as \emph{CoSAMP} \cite{
needell2009cosamp} and hard-thresholding approaches, e.g.  IHT \cite{BlumensathDavies:2008:IterativeThresholding} or HTP \cite{Foucart:2011}, 
one modifies to projection step to project onto the union of sub-spaces. 
The efficiency in time and space complexity of hard-thresholding algorithms for recovering structured signals depends on the existence of an efficient projection onto the structure set. 
Notable exceptions include the relaxiation to approximate projections~\cite{BahEtAl:2019,HegdeIndykSchmidt:2015:Approximation}.

Hierarchical sparsity can be regarded as the combination of two prominent signal structures:  
One of the earliest examples of model-based compressed sensing are \emph{block-sparse} signals \citep{EldarKuppinger2010,EldarMishali:2009b,EldarMishali:2009a,stojnic2009reconstruction}. 
Block-sparse signals are thereby blocked signals $(\vec{x}_1, \dots, \vec{x}_N)$ where only a few $\vec{x}_i$, say $s$, are assumed to be non-zero, but not necessarily in themselves sparse. In other words, they are $(s,(n_1, \dots, n_N))$-sparse. 
The second special case is \emph{level-sparsity} \citep{AdcockEtal:2013,LiAdcock:2016} which are block vectors $(\vec{x}_1, \dots, \vec{x}_N)$, where a sparsity $\sigma_i$ is assumed for each block. Thus, level-sparse signals can be viewed as special cases of hierarchical, namely $(N, \sigmas)$-sparse, signals. 

Original work on block-sparsity, focused on convex recovery algorithms, which can be understood as using the atomic norm \citep{chandrasekaran2012convex} associated with the group sparse vectors as a regularizer, the so called $\ell_{1,2}$-norm,
$\norm{\vec{x}}_{1,2} = \sum_{k=1}^N \norm{\vec{x}}_2$.
A greedy algorithm for block-sparse vectors is proposed in
Ref.\ \cite{MajumdarWard:2009}. 
The atomic norm of hierarchically sparse vectors is the $\ell_1$-norm -- hence, the analogous strategy does not directly carry over. 
Instead, hierarchical sparse vectors were introduced in a line of work
\citep{FriedmanEtAl:2010, SimonEtAl:2013,SprechmannEtAl:2010, SprechmannEtAl:2011} that uses a convex combination of the $\ell_1$ and $\ell_{1,2}$ as a regularizer, giving rise to the 
\emph{HiLasso algorithm}, 
a hierarchical soft-thresholding algorithm.  HiLasso has been equipped with recovery guarantees based on notions of coherence. 
There are also non-convex generalizations of this approach \cite{Wohlberg2013}. 
Further generalizations of the convex combinations of norms for structured sparse signals can be found in Refs.\  \cite{BachEtAl:2012,JenattonEtAl:2011a}.
The \emph{orthogonal matching pursuit algorithm} has been generalized to hierarchical sparse vectors in Ref.\ \cite{LiuSun:2011}. 
We here build on generalizations of the hard-thersholding algorithms employing the efficient projections onto hierarchically sparse vectors \cite{HiHTP,RothEtAl:iTwist:2016}. }

Within model-based compressed sensing recovery guarantees for the generalized algorithmic approaches can be transferred using the aforementioned generalized RIP~\cite{BlumensathDavies:2009:SamplingTheorems} restricted to the signal structure. 
(See also Refs.~\cite{JungeLee:2020:Generalized,TraonmilinGribonval:2018:Stable} for further generalizations.) %
In particular, for Gaussian random matrices one directly derives sampling complexities scaling with the logarithms of the cardinality of the union.  
Going beyond such improvements of polylog-factors in the sampling complexity,  certain measurement ensembles that feature more structure can exhibit a structured version of RIP while not being amenable to standard RIP analysis. 
This is for example a core motivation for level-sparsity \cite{ AdcockEtAl:2021:Benefits,BastounisHasen:2017:OnTheAbsence}.  
For hierarchical sparsity of particular importance, HiRIP can be established for Kronecker product measurement \cite{roth2018hierarchical,RothEtAl:iTwist:2016} without requiring each factor to have the corresponding RIP itself as in unstructured Kronecker compressed sensing \cite{DuarteBaraniuk:2012,JokarMehrmann:2009}. 
The hierarchical measurement operator studied here are a considerable generalization of Kronecker product measurements for which comparable guarantees can be established. 
Compared results for block sparse or level-sparse signals, the analysis of hierarchical measurement operator crucially relies on the interplay between sparsity assumption on different hierarchy levels.

\paragraph{Outline}
The remainder of the work outline is as follows: The main results are presented and discussed in Sections~\ref{sec:HiRIP} and \ref{sec:blocks}, respectively. Section \ref{sec:numerics} is dedicated to numerical experiments. In particular, two applications are introduced and discussed.  Most proofs are postponed to Section \ref{sec:proofs}.

\paragraph{Notation}
For $p \in \N$, we denote by $[p]$ the set of integers between $1$ and $p$. The $\norm{\cdot }$ always denotes the Euclidean norm of $\K^n$. The expression $f \leqsim g$, where $f$ and $g$ are entities depending on parameters $\pi$ with values in $\mathbb{P}$, means' $f$ is majorized by $g$ up to a multiplicative constant', i.e., that there exists a $C >0$ so that $f_\pi\leq C \cdot g_\pi$ for all $ \pi \in \mathbb{P}$. For a set $S$, $\abs{S}$ denotes its cardinality.

\section{HiRIP-properties of general hierarchical measurement operators} \label{sec:HiRIP}

Let us get straight to the formal statement of the first main result of this work.

 \begin{theo} \label{th:HiRIP}
  	Let $\vec{H}$ be a hierarchical measurement operator, as in \eqref{eq:hierarchical}, and $s$, $\sigmas=(\sigma_1, \dots, \sigma_N)$ hierarchical sparsity levels. Assume that 
  	\begin{itemize}
  	    \item The top-level matrix $\vec{A}$ obeys the $s$-RIP with constant $\delta_s(\vec{A})$.
  	    \item The sub-level matrices $\vec{B}_i$ all obey the $\sigma_i$-\af{(Hi)}RIP with constants $\delta_{\sigma_i}(\vec{B_i})$.
  	\end{itemize}
  	 Then $\vec{H}$ obeys the \af{$(s,\sigmas)$-}HiRIP, with
  	 \begin{align*}
  	 	\delta_{(s,\sigmas)}(\vec{H}) \leq \delta_s(\vec{A}) + \sup_{i} \delta_{\sigma_i}(\vec{B}_i) + \delta_s(\vec{A}) \cdot \sup_{i} \delta_{\sigma_i}(\vec{B}_i).
  	 \end{align*}
\end{theo}

  The intuition of the result is the following: if both the top-level and sub-level matrices possess their `respective' RIPs, the hierarchical measurement operator built from them has the HiRIP. 
  This behaviour does in general \emph{not} manifest itself for the RIP property of the hierarchical operator. 
  This becomes apparent from the special case of Kronecker-product measurements. 
  Indeed, as is shown in Ref.\ \cite{jokar:2009sparse}, we in fact have
  \begin{align*}
      \delta_{s}(\vec{A} \otimes \vec{B)} \geq \max (\delta_s(\vec{A}), \delta_s(\vec{B})).
  \end{align*}
  That is, in order for the Kronecker product $\vec{A} \otimes \vec{B}$ to have the $s$-RIP, both $\vec{A}$ and $\vec{B}$ needs to have it. Note further that $(s,\sigma)$-sparse signals are not $s$-sparse, but rather $s\sigma$-sparse. 
  For a discussion of implications in the context of MIMO, 
  see also Ref.\  \cite{ShabaraKoksalEkici:2021}.
  It is thus safe to say that Theorem \ref{th:HiRIP} implies that the hierarchical sparsity framework opens up for recovery for a class of signals and measurement that cannot be treated by standard compressed sensing.

  Theorem~\ref{th:HiRIP} is a significant generalization of an analogous statement for Kronecker product measurements derived in Ref.\ \citep{roth2018hierarchical} and   
  the proof here is actually quite different compared to the one in Ref.\ \citep{roth2018hierarchical}. 
  \af{%
  The main new idea is to 
  use that if $\vec{A}$ acts isometrically on $s$-sparse vectors, it induces an  isometric action on \emph{bi-sparse} matrices. 
  \begin{defi}
  A square matrix $\vec X \in \K^{N,N}$ is $s$-bisparse
  if there exists $S \sse [N]$ with $\abs{S}\leq s$ so that
  	\begin{align*}
  		X_{i,j} =0 \quad  \text{ if $i$ or $j$ is outside $S$.}
  	\end{align*}
 \end{defi}
  The following lemma now captures how $\vec A$ acting simultaneously on the row and column space of an Hermitian matrix distorts its nuclear norm. 
To this end, let $\norm{\vec{X}}_*$ denote the nuclear norm of $\vec{X}$, i.e., the sum of the eigenvalues of $\vec{X}$ and \af{ $\sprod{\, \cdot, \, \cdot}$ denote the Hilbert-Schmidt inner product,  $\sprod{\vec{A},\vec{B}} = \tr(\vec{A}^*\vec{B})$}.
}  
\begin{lem} \label{lem:SchattenConc}
  	Let $\vec{A} \in \K^{M,N}$ obey the $s$-RIP. Assume that  $\vec{X} \in \K^{N,N}$ is an $s$-bisparse Hermitian matrix. 
    It holds that
  	\begin{align*}
  		\abs{\sprod{\vec{A}^*\vec{A},\vec{X}} - \norm{\vec{X}}_* }\leq \delta_s(\vec{A}) \norm{\vec{X}}_*.
  	\end{align*}
  \end{lem}
 
  \begin{proof}
  		 Since $\vec{X}$ is Hermitian, we may decompose it as follows
  		 \begin{align*}
  		 	\vec{X} = \sum_{i=1}^N \lambda_i \vec{x}_i \vec{x}_i^*.
  		 \end{align*}
  		 Here, $\vec{x}_i$ are normalized eigenvectors of $\vec{X}$ and $\lambda_i$ are the eigenvalues of $\vec{X}$. Since $\vec{X}$ is $s$-bisparse, there exists a set $S \sse [N]$ and $\abs{S}\leq s$ such that $\supp (\vec{x}_i) \sse S$ whenever $\lambda_i \neq 0$. We therefore get
  		 \begin{align*}
  		 	\sprod{\vec{A}^*\vec{A},\vec{X}} = \sum_{i=1}^N \lambda_i \sprod{\vec{A}^*\vec{A},\vec{x}_i \vec{x}_i^*} = \sum_{i=1}^N \lambda_i \sprod{\vec{A}\vec{x}_i, \vec{A} \vec{x}_i}.
  		 \end{align*}
  		 Since $\vec{A}$ has the $s$-RIP and all $\vec{x}_i$ are $s$-sparse and normalized, we get
  		 \begin{align*}
  		 	1- \delta_s(\vec{A}) \leq \sprod{\vec{A}\vec{x}_i, \vec{A} \vec{x}_i} \leq 1+ \delta_s(\vec{A}),
  		 \end{align*}
  		 and therefore
  		 \begin{align*}
  		 	 \abs{\sum_{i=1}^N \lambda_i (\sprod{\vec{A}\vec{x}_i, \vec{A} \vec{x}_i}-1) }\leq \sum_{i=1}^N \abs{\lambda_i} \delta_s(\vec{A}).
  		 \end{align*}
  		 The claim follows.
  \end{proof}
  
  We may now prove the theorem.
  
   \begin{proof}[Proof of Theorem \ref{th:HiRIP}]
  		Let $\vec{x} = (\vec{x}_1, \dots , \vec{x}_N)$ be an $(s, \vec{\sigma})$-sparse vector. Let $S$ denote the block support of $\vec{x}$, i.e.,
  		\begin{align*}
  			\vec{x}_i =0  \quad \text{ for }i \notin S.
  		\end{align*}
  		We have
  		\begin{align*}
  			\norm{\vec{H}(\vec{x})}^2 &= \sprod{\sum_{i=1}^N \vec{a}_i \otimes (\vec{B}_i\vec{x}_i), \sum_{j=1}^N \vec{a}_j \otimes (\vec{B}_j \vec{x}_j)} = \sum_{i,j=1}^N \sprod{\vec{a}_i, \vec{a}_j} \sprod{\vec{B}_i \vec{x}_i, \vec{B}_j \vec{x}_j} = \sprod{\vec{A}^*\vec{A}, \vec{G}},
  		\end{align*}
  		where we have defined the matrix $\vec{G} \in \K^{N,N}$ through
  		\begin{align*}
  			G_{i,j} = \sprod{\vec{B}_i \vec{x}_i, \vec{B}_j \vec{x}_j}.
  		\end{align*}
  		This matrix is Hermitian and also $s$-bisparse, since $G_{i,j} =0$ when $i$ or $j$ is not in $S$. Lemma \ref{lem:SchattenConc} therefore implies
  		\begin{align*}
  			\abs{\sprod{\vec{A}^* \vec{A}, \vec{G}} - \norm{\vec{G}}_*}\leq \delta_s(\vec{A}) \norm{\vec{G}}_*.
  		\end{align*}
  		Now notice that $\vec{G}$ can be written as $\vec{G}=\vec{M}^*\vec{M}$,
  		where $\vec{M} \in \C^{m,N}$ is defined through
  		\begin{align*}
  			\vec{M} \vec{c} = \sum_{i=1}^N c_i \vec{B}_i \vec{x}_i.
  		\end{align*}
  		We have $\norm{\vec{G}}_*= \norm{\vec{M}}_F^2=\sum_{i=1}^N \norm{\vec{B}_i \vec{x}_i}^2 $. Since each vector $\vec{x}_i$ is $\sigma_i$-sparse, we get 
  			\begin{align*}
  				\sum_{i=1}^N (1-\delta_{\sigma_i}(\vec{B}_i))\norm{\vec{x}_i}^2  &\leq \sum_{i=1}^N \norm{\vec{B}_i \vec{x}_i}^2 \leq \sum_{i=1}^N (1+\delta_{\sigma_i}(\vec{B}_i))\norm{\vec{x}_i}^2 .
  			\end{align*}
  			This implies
  			\begin{align*}
  		        \abs{\norm{\vec{H}(\vec{x})}^2 - \norm{\vec{x}}^2} \leq \abs{\sprod{\vec{A}^* \vec{A}, \vec{G}} - \norm{\vec{G}}_*} &+ \abs{\norm{\vec{G}}_* - \norm{\vec{x}}^2} \leq \delta_s(\vec{A}) \norm{\vec{G}}_*  + \sup_{i} \delta_{\sigma_i}(\vec{B}_i) \norm{\vec{x}}^2 \\
  				&\leq (\delta_s(\vec{A}) +\sup_{i} \delta_{\sigma_i}(\vec{B}_i)  + \delta_s(\vec{A}) \sup_{i} \delta_{\sigma_i}(\vec{B}_i))\norm{\vec{x}}^2 \, .
  			\end{align*}
  \end{proof}

To exemplify its practicality, let us discuss the result within the IoT scenario outlined in the introduction.  
The theorem's statement can be directly translated to the model: if
\begin{itemize}
    \item each code book $\vec{B}_i$ has the $\sigma_i$-RIP,
    \item the matrix $\vec{A}$ of channel gains has the $s$-RIP,
\end{itemize} 
the hierarchical `base station operator' $\vec{H}$ will have the $(s, \vec \sigma)$-HiRIP. 
This in turn implies that the base station can use the hierarchical recovery algorithms to recover all the users messages. 
The two assumptions are fulfilled in several practical settings: First, the users can use a standard ensemble of compressed sensing matrices, such as random Gaussian matrices $\vec{B}_i$ or sub-sampled bounded orthogonal systems as code books.  
For such ensembles it is well-known that a random matrix has the $\sigma_i$-RIP with high probability if $m \geqsim \sigma_i \operatorname{polylog}(n_i)$  \cite{FouRau13}. 

As for the channel gain matrix, the situation depends on the geometry of the receiving antennas. Let us discuss two important special cases: 
If the antennas are well-separated, their channel gains can be modelled as random Gaussians. 
If the antennas are instead arranged in a uniform linear array, we can model $\vec{A}$ as a Fourier matrix \cite{Wunder_TWC19}.
Again by classic results \cite{FouRau13}, we may randomly sub-sample $M \sim s \log(N)^4$ of the rows of a Fourier matrix and still end up with a matrix with an $s$-RIP. 
\af{In fact, the latter random subsampling technique is the standard argument for applying compressed sensing in (hybrid) RF beamforming techniques for MIMO networks. The main advantage is that the RF beamforming network operates at a drastic dimensionality reduction from $N$ antennas to just $M$ RF ports. In fact any massive MIMO array (say with  more than $100$ antennas) that would require an amount of RF ports in the order of the number of antennas is infeasible in practice both in terms of hardware (e.g. expensive RF components), energy and computational costs (analog/digital converters.). For further details see \cite{Wunder_TWC19}.}

\af{Overall, we conclude that the base station needs no more than
 \begin{align*}
    M \cdot m \geqsim s\max_i \sigma_i \cdot \mathrm{polylog}(N,n)
 \end{align*}
 total measurements to allow for recovery of all $(s, \vec{\sigma})$-sparse vectors.
 This is the same scaling as required for establishing HiRIP for unstructured  Gaussian measurement operators  (up to log-factors) \citep{HiHTP}. }

 \begin{rem}
 These considerations naturally extend the results derived for Kronecker operators from Ref.\ \citep{RothEtAl:iTwist:2016} to a considerably more general measurement operators that use varying sparsity levels, ambient dimensions and sub-level matrices $\vec{B}_i$. 
 The coding scenario explained above already exemplifies an instances where such a general model is practically required. Here it is desired that each user can use his/her own private code book.  
 \end{rem}

\subsection{Necessity of RIP of the top- and sub-level matrices}
It is interesting to ask to which extent the conditions of Theorem \ref{th:HiRIP} are necessary. 
In particular: Can the $(s,\sigmas)$-HiRIP-constants of $\vec{H}$ be smaller than what the RIP properties of the constituent matrices suggest via Theorem \ref{th:HiRIP}? 
In this section, as well as in the subsequent one, we derive results that answer this question.  
Let us begin by discussing the sub-level matrices. Note that if $\vec H$ has the $(s, \sigmas)$-RIP, each matrix
\begin{align*}
	\norm{\vec{a}_i}_2 \vec{B}_i
\end{align*}
must have the $\sigma_i$-RIP. To see this, let $i$ be arbitrary, $\vec{x}_i$ a $\sigma_i$-sparse vector, and $\widehat{\vec{x}}_i=(0, \dots, \vec{x}_i, \dots,0)$ the vector with its $i$:th block equal to $\vec{x}_i$, and zeros elsewhere. We then have
\begin{align*}
	\abs{ \norm{\vec{a}_i}^2 \norm{\vec{B}_i\vec{x}_i}^2 - \norm{\vec{x}_i}^2} &= \abs{\norm{\vec{H}(\widehat{\vec{x}}_i)}^2 - \norm{\widehat{\vec{x}}_i}^2} \leq \delta_{(s,\sigmas)}(\vec{H}) \norm{\vec{x}_i}^2,
\end{align*}
since $\widehat{\vec{x}}_i$ is $(1, \sigma_i)$-sparse. We can therefore conclude the following result.

\begin{prop} \label{prop:negres}
    Assume that $\delta_{(s,\sigmas)}(\vec{H}) <1$, and that each column $\vec{a}_i$  in $\vec{A}$ is normalized. Then
    \begin{align*}
       \delta_{\sigma_i}(\vec{B}_i) \leq \delta_{(s,\sigmas)}(\vec{H}).
    \end{align*}
\end{prop}

\begin{rem}
    The assumption of normalized columns is not really a restriction. Indeed, since $\vec{a}_i \otimes \vec{B_i}= (\lambda \vec{a}_i) \otimes (\lambda^{-1} \vec{B}_i)$ for each $\lambda>0$, we may always simultaneously re-scale the columns and sub-level matrices to  $\Vert \vec{a}_i \Vert=1$ for each $i$, while keeping the hierarchical operator constant.
\end{rem}

As for the RIP-properties of the top-level matrix $\vec{A}$, the situation is  more complicated.  It is in particular not necessary for $\vec{A}$ to have the $s$-RIP. After all, the $\vec{B}_i$  can map into $N$ pairwise orthogonal sub-spaces $V_i$. Then we have
\begin{align*}
    \left\|\sum_{i=1}^N \vec{a}_i \otimes \vec{B}_i\vec{x}_i\right\|^2 = \sum_{i=1}^N \norm{\vec{a}_i}^2\norm{\vec{B}_i\vec{x}_i}^2.
\end{align*}
Thus, if each matrix has the $\sigma_i$-RIP, $\vec{H}$ will have the $(s,\sigmas)$-HiRIP already when each column of $\vec{A}$ is normalized. In particular, they may be equal, so that $\vec{A}$ does not have the $s$-RIP for any $s>1$. 

In the above example, the block operators $\vec{B}_i$ already allow by themselves for a de-mixing of their inputs. 
If this is not the case, e.g., if $m$ is impermissively small, we need $\vec{A}$ to ``help them'' via having acting isometrically itself. 
To be concrete, we have the following result. The proof for it is \af{conceptually} simple but technical, whence we postpone it to Section~\ref{sec:proofs}.

\begin{prop} Let $\vec{H}$ have the $(s,\sigmas)$-HiRIP. Assume that  for each subset $S \sse [N]$ there exist \af{$\sigma_i$-sparse vectors $\vec{g}_i$, $i\in S$  so that
\begin{align*}
\vec{B}_i \vec{g}_i = \vec B_j \vec{g}_j, \quad i,j \in S.
\end{align*}} 
Then 
\begin{align*}
	\delta_s(\vec{A}) \leq \frac{\delta_{s, \sigmas}(\vec{H}) + \sup_{i} \delta_{\sigma_i}(\vec{B}_i)}{1-\sup_{i} \delta_{\sigma_i}(\vec{B}_i)} \,.
\end{align*} \label{prop:optimality}

\end{prop}

The aforementioned discussion focused on two 
extreme cases: 
In one case, $\vec{B}_i$ map into pairwise orthogonal sub-spaces.
In the other case, each subset of $s$ matrices $\vec{B}_i$ maps sparse vector into the same sub-space of $\K^m$.
We now take a look at the realm in between the two extreme cases and ask what can be gained already if the $\vec{B}_i$ map sparse vectors into sufficiently `incoherent' sub-spaces of $\K^m$?

\section{Incoherent sub-level matrices} \label{sec:blocks}

In this section, we establish a refined version of Theorem~\ref{th:HiRIP} for the case when the sub-level matrices are incoherent. Let us make the latter notion precise.
\begin{defi}[Pairwise $(\delta,\sigmas)$-incoherence]
 Let $\delta>0$. We say that the collection of operators $\vec{B}_i \in \K^{m\times n_i}, i \in [N]$ are \emph{pairwise $(\delta,\sigmas)$-incoherent} if for each $i\neq j$,
  \begin{align*}
      \abs{\sprod{\vec{B}_i \vec{v}_i ,\vec{B}_j \vec{v}_j}}\leq \delta.
  \end{align*}
  for each pair of normalized and $\sigma_i$- and $\sigma_j$-sparse, respectively, vectors $\vec{v}_i$ and $\vec{v}_j$.
\end{defi}
Note that the above condition can be also formulated in terms of the sparse block-coherence introduced in Ref.\  \citep{SprechmannEtAl:2011}.
Collections of incoherent operators are not hard to construct. In fact, we obtain such a collection by \emph{independently} sampling from several common distributions, as summarized in the following proposition.

\begin{prop} \label{prop:incoherentCollections}
        \noindent  1. Let each entry of each sub-level matrix $\vec{B}_i \in \R^{m,n}$ be independently drawn from a sub-Gaussian distribution $\mathscr{D}$ \cite[sec. 7.4]{FouRau13}.
        Suppose that for some $\sigma\in \N$,
        \begin{align*}
            m \geqsim \delta^{-2}\left(\sigma \log\left(\frac{n}{\sigma}\right) + \log(N)\right),
        \end{align*}
        with implicit constants only depending on the sub-Gaussian parameters of $\mathscr{D}$. 
        Then the collection is $(\delta,\sigma)$-incoherent with high probability.
        
       \noindent 2. Let $\vec{U} \in \K^{n,n}$ be a unitary matrix with bounded entries such that
        \begin{align*}
            \sqrt{n}\sup_{k,\ell}  \abs{U_{k,\ell}} \leq K
        \end{align*}
        for some constant $K \in \R$.
        For each $i$, let the matrix $\vec{B}_i \in \K^{m,n}$ be constructed by 
        \begin{enumerate}
            \item uniformly and independently sampling $m$ rows of $\vec{U}$
            \item  multiplying each of the rows with a uniform random sign
            \item rescaling the rows by a factor $m^{-1/2}$.
        \end{enumerate} 
        We assume that $\vec{B}_i$ is independent of $\vec{B}_j$, for $i \neq j$. 
        Suppose for some $\sigma \in \N$ that
        \begin{align*}
            m \geq C K^2\sigma\delta^{-2} \log(n)^4\log(N),
        \end{align*}
         the collection is $(\delta,\sigma)$-incoherent with probability higher than $1-n^{-\log{n}^3}$.
\end{prop}

\af{\begin{rem}
Note that in this result, the sparsity $\sigma$ is scalar. The first part could probably be generalized to more sparsity levels. However, the second cannot. The reason for this is that it is not hard to give conditions under which sub-Gaussian matrices have the $\sigma$-RIP for $\sigma$ of arbitrarily many levels -- the same is however not true for the subsampled unitary matrices. 
\end{rem}}

The proof of this proposition, which employs standard compressed sensing techniques, is postponed to Section \ref{sec:proofs}.

Intuitively, if  the sub-level matrices $(\vec{B}_i)_{i\in[N]}$ form  a pairwise incoherent collection, they can to some extent already intrinsically separate the contributions of the individual blocks of a hierarchically sparse vector from the `single-shot' measurement $\sum_{i=1}^N \vec{B}_i \vec{x}_i$. 
Therefore, the top-level matrix $\vec{A}$ is expected to not be required to `help them'  by having the $s$-RIP to the same extent as in the general setting. This is indeed the case. A formal result is as follows.

\begin{theo}\label{th:Incoherent Blocks}
    Let $\vec{B}_i \in \K^{m\times n_i}$ for $i\in[N]$, be a pairwise $(\delta_{2\sigma}^*,\af{\sigmas})$-incoherent family. Further assume that
    \begin{align*}
        \sup_{i} \delta_{\sigma_i}(\vec{B}_i) \leq \delta_\sigma^*
    \end{align*}
    Let further $\vec{H}$ be a hierarchical measurement operator formed by the $\vec{B}_i$ and a matrix $\vec{A}\in \K^{M\times N}$ with $\delta_{2s}(\vec{A})<1$ for some $s\in\N$. Then,  for any $\lambda\in\N$, $\vec{H}$ has the $\hat{\vec{s}}=(\lambda t, \af{\sigmas})$-HiRIP with
    \begin{align*}
        \delta_{\hat{\vec{s}}}(\vec{H}) \leq \delta_{s}(\vec{A}) + \delta_{\sigma}^* + \delta_s(\vec{A}) \delta_\sigma^* + \lambda\sqrt{s} \delta_{2s}(\vec{A}) \delta_{2\sigma}^*
    \end{align*}
\end{theo}

The proof employs similar ideas as the one of Theorem \ref{th:HiRIP}, but is more technically involved. We therefore postpone it in its entirety to Section \ref{sec:proofs}. Let us here instead stress its intuitive meaning:  A small value of  $\delta_{2\sigma}^*$ (that is, a highly incoherent sub-level collection),  can be `traded in' for a higher value of $s$ in the hisparsity-index $(s, \sigmas)$. 

Note that the parameters $s$ and $\lambda$ can be chosen in various ways to yield the same value of $s\lambda$. 
This arguably makes the result slightly hard to decipher. 
In particular, given values of $\delta_{2\sigma}^*$ and $s$, it does not clearly state the requirements on the top-level matrix to obtain a $(t,\vec{\sigma})$-HiRIP. 
The following result sheds more light on the situation when the  top-level matrix is Gaussian. Again, the proof is postponed to the Section \ref{sec:proofs}.

\begin{prop}
\label{prop:s_vs_delta}
    Assume that $(\vec{B}_i)_i$ is as in Theorem \ref{th:Incoherent Blocks}, and that $\vec{A}\in \K^{M\times N}$ is a Gaussian matrix. Let furthermore $\delta,\epsilon>0$. Provided
    \begin{align*}
        M \geqsim \frac{(t\delta_{2\sigma}^*)^2}{\delta^2}\log\left(\frac{N(1+\delta_\sigma^*)^2}{(t\delta_{2\sigma}^*)^2}\right) + \log(\epsilon^{-1}),
    \end{align*}
    the hierarchical measurement operator $\vec{H}$ defined by $\vec{A}$ and $(\vec{B}_i)_i$ obeys
    \begin{align*}
        \delta_{(t,\sigmas)}(\vec{H}) \leq \delta + \delta_{\sigma}^*
    \end{align*}
    with a probability at least $1-\epsilon$.
\end{prop}

The above result states that in the case of a Gaussian $\vec{A} \in \K^{M,N}$, we need $M$ to be of the order $(s\delta_{2\sigma}^*)^2$ to allow for $s$-sparse signals on the `block level', rather than $s$. For small $\delta_{2\sigma}^*$, we may hence obtain the RIP already when $M$ is less than $s$ -- a behaviour which cannot be explained by Theorem~\ref{th:HiRIP}.

\begin{rem}The quadratic dependence on $s\delta_{2\sigma}^*$ here is of course not sample-optimal. We actually believe that it is likely to be an artefact of the proof. We leave the possible strengthening of the result to future work.\end{rem}

\section{Numerical simulations} \label{sec:numerics}

In this section, we perform numerical simulations to showcase the practical relevance of our results. First, we empirically verify the implications of our main results, Theorem~\ref{th:HiRIP} and \ref{th:Incoherent Blocks}. Then, we explore two potential applications from mobile communications.
In our experiments, we apply the 
\emph{hierarchical hard thresholding pursuit (HiHTP)} algorithm. 
HiHTP is a low-complexity algorithm for solving hierarchically sparse compressed sensing problems of the form
\begin{equation}
\label{eq:problem}
    \min\limits_x \frac12\|y-\vec{H}x\|^2 \quad \text{ subject to $x$ is $(s,\sigmas)$-sparse,} 
\end{equation}
where $\vec{H}$ is a linear operator. 
If $\vec{H}$ has a suitable HiRIP for an appropriate sparsity level, HiHTP is theoretically guaranteed to robustly recover any $(s,\sigmas)$-sparse vector. The interested reader is referred to
Ref.\ \cite{HiHTP} for more details.

The algorithm in each iteration performs a gradient descent step and projects the result onto $(s,\sigma)$-sparse vectors to obtain a support estimate.  
Subsequently, the new iterate is obtained by solving the linear least-square problem restricted to the support estimate. 
The algorithm terminates once the support of two successive iterates does not change or another suitable stopping criterion is reached. 
Importantly, the projection $T_{s,\sigmas}$ onto the set of $(s,\sigmas)$-sparse signal (line 3 of Algorithm \ref{alg:hihtp}) can be performed via hierarchical hard-thresholding in time complexity $O(Nn)$. We again refer to Ref.\  \cite{HiHTP} for implementation details.

\begin{algorithm}[tb]

\SetAlgoLined
\SetKwInOut{Input}{input}\SetKwInOut{Output}{output}
\SetKwInOut{Init}{initialize}\SetKwFunction{Break}{break}
\DontPrintSemicolon
\Input{Problem data $\vec{y}\in\K^m$, $\vec{H}\in\K^{m\times Nn}$, hisparsity $(s,\sigmas)$}
\Init{$\vec{x}^{(0)} = 0$}
\Repeat{stopping criterion is met at $t = t^\ast$}{
$\bar{\vec x}^{(t)} = \vec x^{(t-1)} + 
{\vec H}^*\left(\vec y- \vec H {\vec x}^{(t-1)}\right)$; \;
$I^{(t)} = \operatorname{support}\ T_{s,\sigmas}\left(\bar{{\vec x}}^{(t)} \right)$; \;
${\vec x}^{(t)} = \argmin\limits_{{\vec x}} \frac12 \|{\vec y}-{\vec H}{\vec x}\|^2 \quad$ subject to $\quad \supp(x)\subseteq I^{(t)}$; 
}
\Output{$(s,\sigma)$-sparse vector $\vec x^{(t^\ast)}$}
\caption{HiHTP  \label{alg:hihtp}}
\end{algorithm}

\subsection{Verification of Theorem \ref{th:HiRIP} and Theorem \ref{th:Incoherent Blocks}:} 

In a nutshell, Theorem~\ref{th:HiRIP} tells us that HiHTP successfully recovers $(s,\sigmas)$-sparse signals reliably from a hierarchical measurement if both $\delta_s(\vec{A})$ and each $\delta_{\sigma_i}(\vec{B_i})$ are small. It is common wisdom that the latter is achieved with high-probability with suitable drawn random matrices $\vec{A} \in \K^{M,N}$ and $\vec{B}_i \in \K^{m,n}$ if and only if $M \geqsim s\text{ and } m \geqsim \max_i \sigma_i$, up to logarithmic terms in $N$ and $n_i$. Thus, if we fix $M,m,n$ and $N$, HiHTP should be able to recover $(s,\sigmas)$-sparse  for $s\leq s^*$ and $\sigma \leq \sigma^*$ for some thresholds $s^*$ and $\sigma^*$.  Conversely, we expect the recovery to start to fail below thresholds with an identical scaling. 

\begin{figure}

\centering   
\includegraphics[width=.4\textwidth]{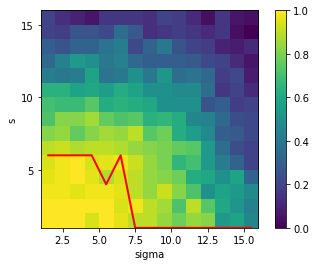}\includegraphics[width=.4\textwidth]{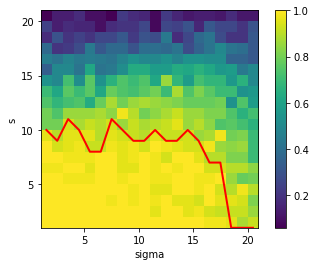}
    \caption{Results for the `equal blocks, equal $\vec{B}_i$'-experiments for $M=35$, $m=45$, $N=40$, $n=100$ (left) and $M=50$, $m=75$, $N=80$, $n=200$ (right). The red line defines a threshold: below it, the recovery probability is at least $94\%$. \label{fig:eqeq}}
\end{figure}

To test whether we observe such a behaviour also in the average performance of the algorithm, we set up an experiment as follows: For each $s$ and $\sigma$ within a range, we draw the top-level matrix $\vec{A}\in\C^{M\times N}$ as a (correctly re-scaled) complex random Gaussian and set each block operator $\vec{B}_i\in \C^{m\times n}$ equal to a common randomly sub-sampled DFT-matrix $\vec{B} \in \C^{m\times n}$. We generate the ground truth signal $\vec{x}$ by drawing a $\sigma$-sparse support at random from the index set $[n]$, draw the values on that support according to a Rademacher distribution, thereby letting all block $\vec{x}_i$ in $\vec{x}$  be equal. By choosing both the measurement matrices $\vec B_i$ and the signal $\vec{x}_i$ identical for all $i$, respectively, we ensure that there is no sub-level incoherence that\af{, in light of Theorem \ref{th:Incoherent Blocks}, } could `help' in the de-mixing.  Thus, we are especially testing the cases that highly depend on the compressed sensing capabilities of $\vec A$ as well. 
The measurement vector $\vec{b}$ is  set equal to $\vec{H}\vec{x}$ (i.e., we make a noise-free experiment) and let the HiHTP algorithm run.  For each set of parameters, $50$ random trials are conducted. We declare a run successful if after at most $25$ iterations, the output  $\vec{x}^*$ of the algorithm obeys $\norm{\vec{x}^*-\vec{x}}/\norm{\vec{x}}<10^{-7}$. 

The results for the two settings $M=35$, $m=45$, $N=40$, $n=100$ and $M=50$, $m=75$, $N=80$, $n=200$ are depicted in Fig.\  \ref{fig:eqeq}. We have marked a (manually determined) approximate iso-line for a recovery probability larger than  or equal to $94 \% = 47/50$. We observe that the region of high-recovery probability has approximately the rectangular shape in agreement with the scaling suggested by Theorem~\ref{th:HiRIP}.

\begin{figure}
   \centering 
   \includegraphics[width=.4\textwidth]{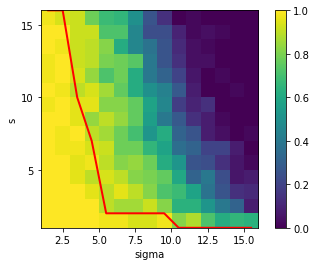}\includegraphics[width=.4\textwidth]{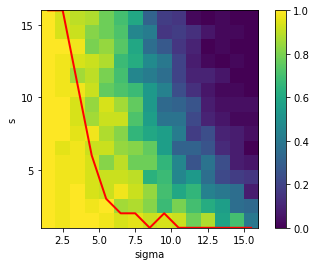}
    \caption{Results for  $M=35$, $m=45$, $N=40$, $n=100$ for the `equal blocks, different $\vec{B_i}$' (left) and `equal $\vec{B}_i$, different blocks' (right) setup. The red line is a manually determined threshold-line for a recovery probability of at least $94\%$. \label{fig:neqneq}}
\end{figure}

To showcase Theorem \ref{th:Incoherent Blocks}, we repeat the experiment for $M=35$, $m=45$, $n=40$, $N=100$. We draw each $\vec{B}_i$ in the same manner as before but now i.i.d.\ for each $i$.\footnote{Note that formally Proposition \ref{prop:incoherentCollections} only holds when each row of the $\vec{B}_i$ is multiplied with with a random  sign. We do not explicitly account for this in our signal model here.} 
In this setting, the theory suggests that $M$ larger than $(s\delta_{2\sigma}^*)^2$ is sufficient (where the square is probably a proof artefact). Thus, choosing $\sigma$ less than $\sigma^*$ and $s \leq s^*/\delta_{2\sigma}^*$ should be sufficient for recovery. The results of the experiments, depicted on the left of Fig.\ \ref{fig:neqneq}, show this behaviour -- notice that the set of values where recovery is probable is considerably stretched in $s$-direction.

Interestingly, when we repeat the experiment for equal $\vec{B}_i$, but draw each block $\vec{x}_i$ independently at random, the same behaviour can be observed, depicted in Fig.\ \ref{fig:neqneq} on the right.  We suspect that the probability of $\vec{H}$ acting close to isometrically on the subset of hisparse signals with incoherent blocks is high, and that this in turn helps the recovery process. The present theory however cannot fully support any of these claims, and we leave the theoretical analysis of this phenomenon to future work.

\subsection{Massive random sparse coding}  

\begin{figure}
    \centering
    \includegraphics[width=.4\textwidth]{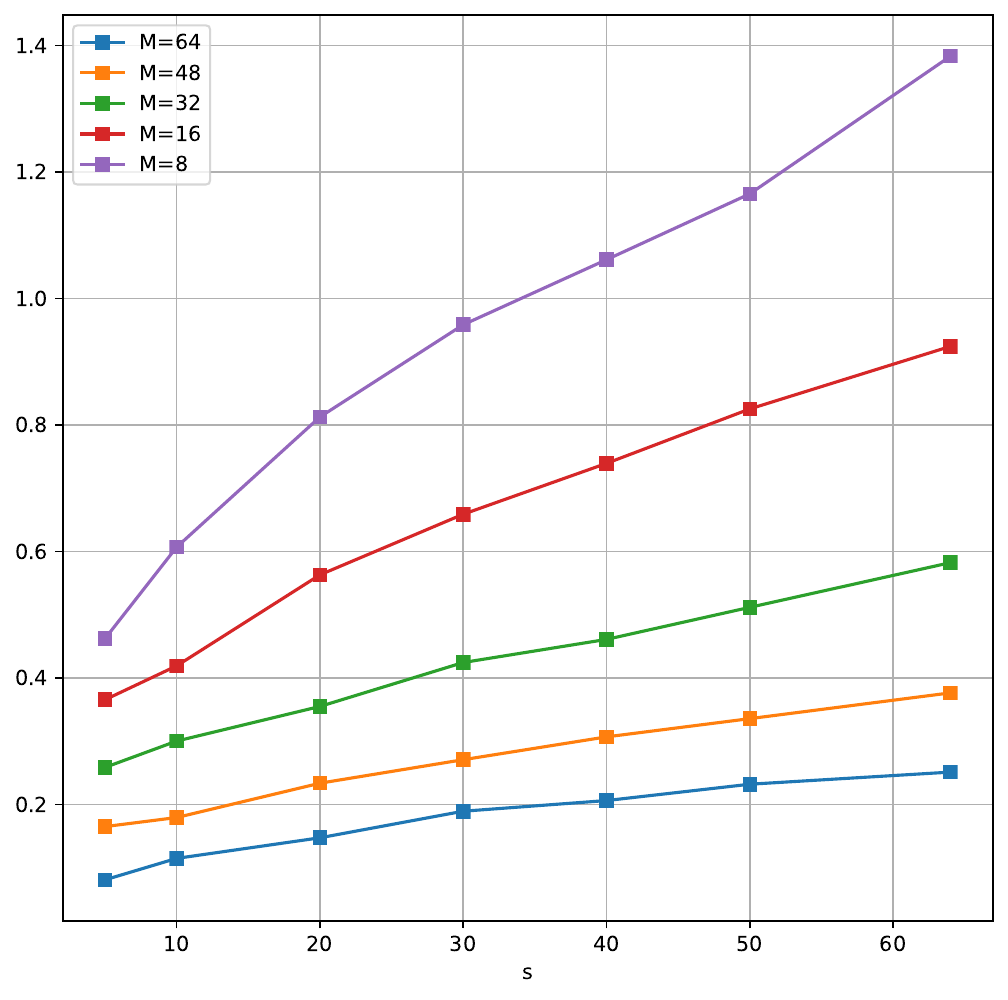}  \includegraphics[width=.4\textwidth]{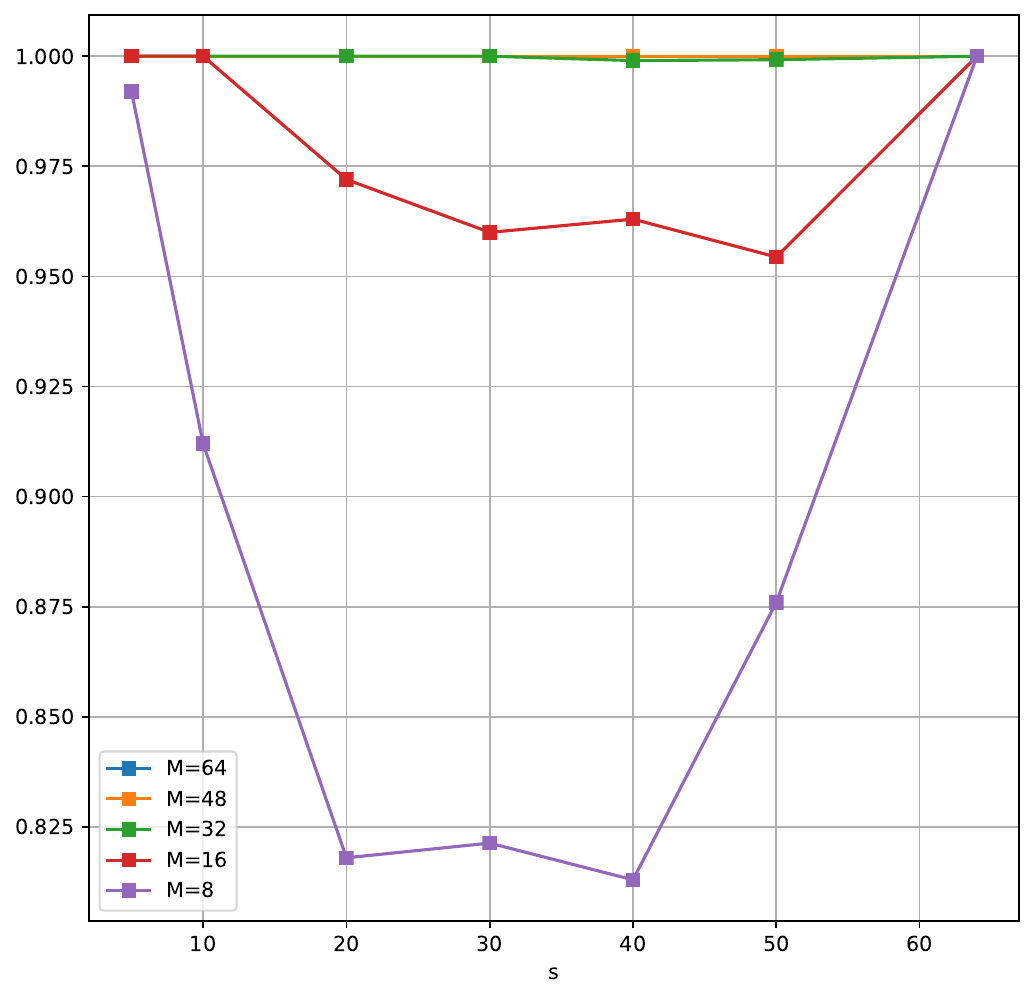}
    \caption{Results for the massive IoT model with varying number of active users $s$ and number of active antennas $M$ for a SNR of 5dB. Left: Relative recovery error. Right: Percentage of correctly detected active users.}
    \label{fig:err_random_coding}
\end{figure}

We move on to an experiment related to the IoT application, namely the model in \eqref{eq:sparseCoding} we outlined in the introduction. The code books $\vec{B}_i \in \C^{m,n}$ are chosen as complex-valued Gaussians, 
 drawn independently for each user. We model $\vec{A} \in \C^{M,N}$ as a randomly sub-sampled Fourier matrix, corresponding to a uniform linear array geometry.  
 The number $M<N$ is the number of randomly sampled antennas.  We vary this number, as well as the number of active users $s$,  while fixing $m=100$, $n=400$, $N=64$, and the sub-level sparsity to $\sigma=10$, synthetically generating data. We then add white Gaussian noise $n\sim \mathcal{N}(0, \eta I)$ to the measurements $y$, where $\eta = 10^{-SNR/10}\|y\|$. We record the mean recovery error $\|\vec{x}-\vec{x}^*\|/\|\vec{x}\|$, as well as the fraction of correctly identified active users, for each configuration over 25 random trials at an (expected) SNR of 5dB. 
 The results are shown in Fig.\ \ref{fig:err_random_coding}. We 
 have observed that for all values of $M$ and $s$, a significant portions of the users are correctly identified as active.  For  $M\geq32$ all active users are found for any $s$. The recovery error also degrades gracefully with a rising number of users.

\subsection{Grouped random access} 

Let us end this section by extending the IoT model towards \emph{random access scenario}.
In such random access, a user that wishes to communicate with the base station chooses a resource at random (out of $\nu$ available resources) and sends a pilot signal associated with that resource. Assuming equal power transmitted from the users, and letting $\vec{b}_{j(i)} \in \C^m,m\leq\nu$, be the pilot signal that user $i$ chooses, the base station
then receives the vector
\begin{align*}
    \vec{y} = \vec{B}\vec{x} = \left(\sum_{i=1}^\nu x_i \vec{b}_{j(i)}\right),
\end{align*}
where $x_i =0$ if user $i$ is inactive, $x_i=1$ if her/he is active, and $\vec{B}=(\vec{b}_1,\dots ,\vec{b}_{\nu})$.
As before, it is reasonable to assume that the user activity is sporadic, which implies that the vector $\vec{x}$ is sparse.
This protocol has a fundamental problem -- if several users choose the same resource, a collision occurs and subsequent communication is impossible since each frequency can only serve one user at a time. The probability of a collision grows fast with the number of users -- a phenomenon commonly referred to as the \emph{birthday paradox}.

\begin{figure}
    \centering
    \includegraphics[width=.75\textwidth]{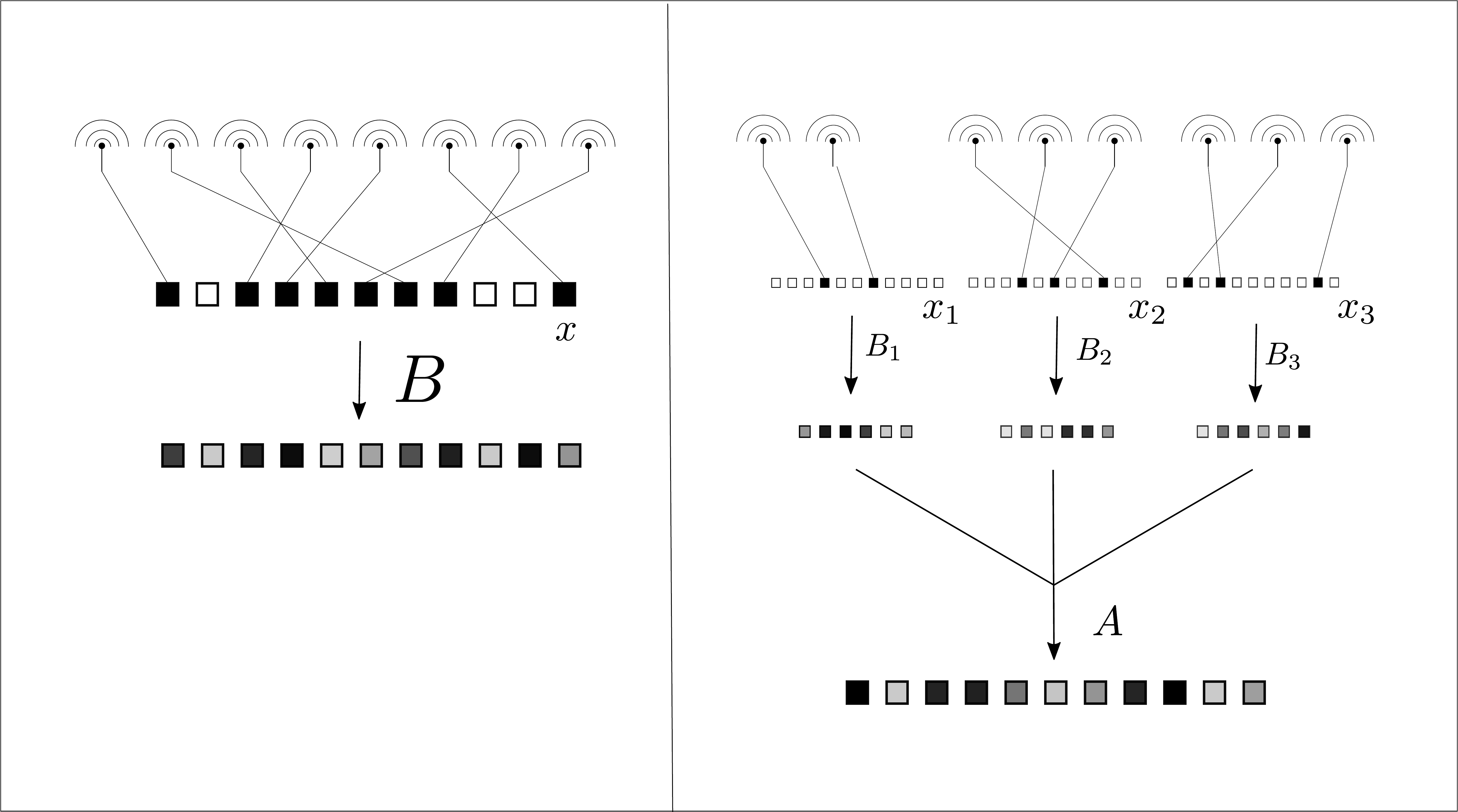}
    \caption{The standard random access (left) vs. the grouped approach proposed here (right).}
    \label{fig:user_detex}
\end{figure}

In order to reduce the probability of collisions, we have proposed to distribute the users in groups \cite{ConferenceVersion_2,Wunder17_Asilomar,Wunder15_Asilomar}. To this end, we randomly subdivide the active users in $N$ groups, and accordingly replace the entries of the vector $x$ with in total $N$ blocks $\vec{x}_i$. We then let the users in each group $i$ choose one of $n$ resources, with $n \leq \nu$. We refer to this protocol as \emph{grouped random access} \cite{ConferenceVersion}. Obviously, there is then \emph{within each block} a much lower probability of collision than before. Furthermore, each block is with high probability sparse, say $\sigma_i$-sparse. Thus, the signal to be recovered, $(\vec{x}_1, \dots, \vec{x}_N) \in (\C^n)^N$ is no longer only sparse, but with high probability  even $(N,\sigmas)$-hierarchically sparse. 

In order for the base station to be able to de-mix the individual block contributions, we may let them send their pilot signals during disjoint time intervals and recover each $\vec{x}_i$ from $\vec{B}_i\vec{x}_i$, for $i=1,\dots ,N$ individually. This would however be very tedious compared to the original, single-shot scheme. To increase efficiency, we may first and foremost let the $\vec{B}_i$ be sub-sampled, i.e. use shorter slots. To still achieve recovery, we propose to mix the contributions $\vec{B}_i \vec{x}_i$ over $M$ incoherent slots (say in time, frequency or space), each time $j$ with a different random modulation $a_{j,i}$. The base station over the course of those slots then receives
\begin{align*}
    \vec{y}_j= \sum_{i=1}^N a_{j,i}\vec{B}_{i} \vec{x}_i, \quad j \in [M],
\end{align*}
which defines a hierarchical measurement. Notably, since all the $N$ blocks may be filled, we arguably are no longer in the  hierarchical compressed sensing realm. However, Theorem \ref{th:Incoherent Blocks} about incoherent blocks still indicates that recovery is possible.

To test the practical performance of the protocol, we perform a simulation on synthetic data. We assume  $\nu=n=512$ available resources and model the measurements at the base station as $\vec{y} = \vec{F}\vec{x}$, where $\vec{F}\in\C^{n\times n}$ is a $n\times n$-DFT matrix. Note that since we are aiming for user detection, we do not necessarily need to recover $\vec{x}$ exactly: we only need to determine which $x_i$ are non-zero. Our baseline method therefore consists of computing $\supp(\vec{F}^{-1}y)$ to obtain the selected resources. 

\begin{figure}
    \centering
    \includegraphics[width=.45\textwidth]{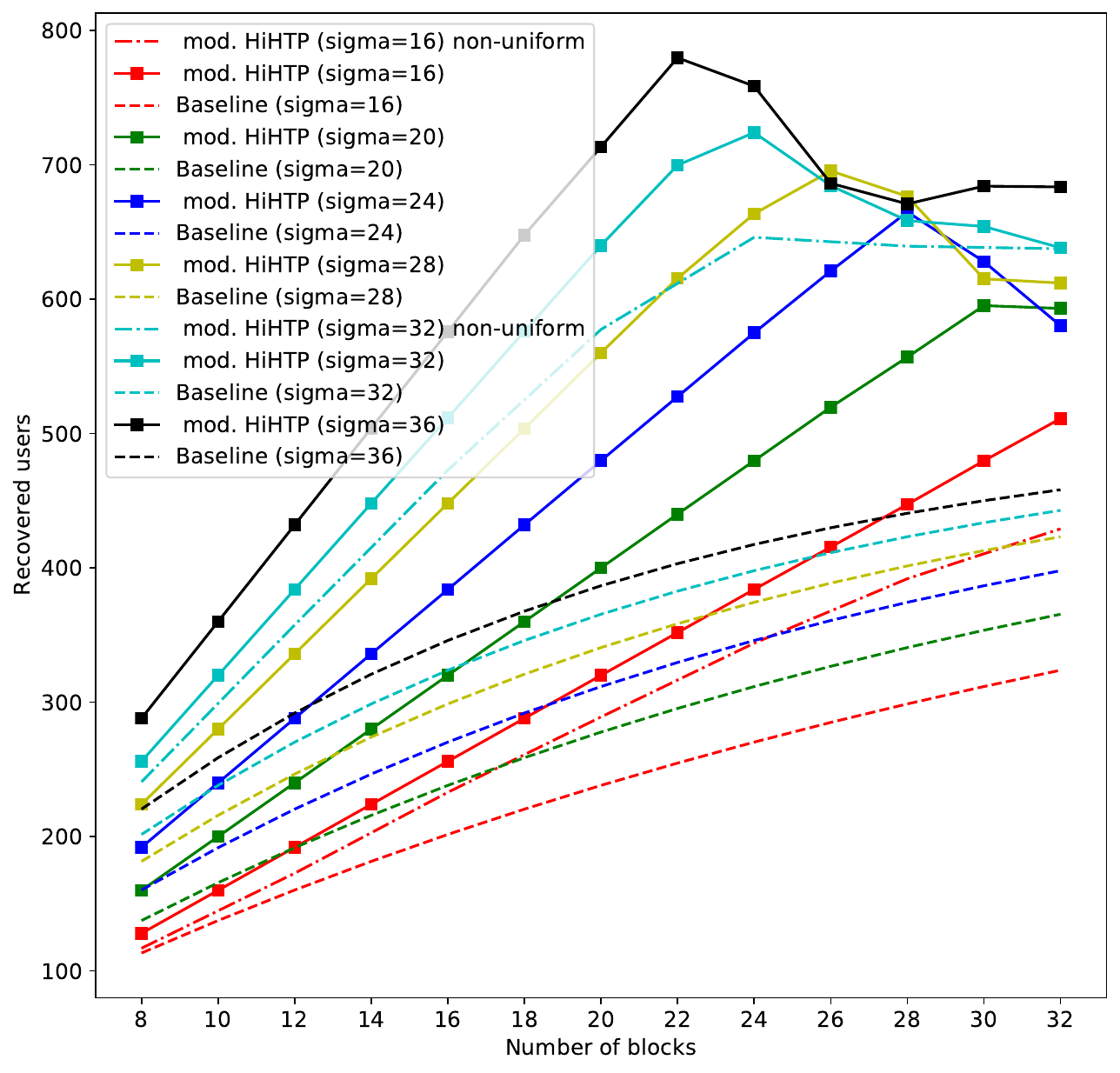}
    \caption{User detection with hierarchical measurements}
    \label{fig:user_detection}
\end{figure}

We compare this to the grouped random access model with $N$ sub-sampled $\vec{B}_i$, which each consist of $m=256$ random rows from the $n\times n$ DFT matrix. The random modulation matrix $\vec{A}\in\C^{M\times N}$, where $M=16$, 
is chosen as complex Gaussian with variance $\tfrac{1}{\sqrt{N}}$. 

 Figure \ref{fig:user_detection} shows the average number of users recovered by HiHTP for varying sparsities $\sigma=16,\dots,36$ and number of blocks $N=8,\dots,32$ over 25 random trials for each configuration. Note that for $N>8$ the system has more pre-image dimensions than measurements, so that we are in a compressed sensing regime. The baseline shown in Figure \ref{fig:user_detection} is computed as the total number of users, $N\cdot \sigma$, minus the expected number of collisions that occur, if these users choose randomly out of the $n=512$ available resources. As can be seen, HiHTP is able to recover many more users compared to the baseline. 
For $\sigma=16$ and $\sigma=32$, we also show the performance in the case that  $\sigma\cdot N$ users are distributed randomly over all available slots (i.e. the block sparsities are not uniformly fixed to $\sigma$). Even in this scenario, where the algorithm operates with the wrong sparsity parameters, reasonable performance is achieved.

  \section{Conclusion} 
  Hierarchical sparsity is a structured notion of sparsity that arises in many applications. 
  The structure also allows for efficient custom-tailored recovery algorithms, such as the HiHTP, but it is more restrictive yielding stronger guarantees than the ones given by classical compressed sensing. 
  In this work, we have introduced a large class of linear operators that are aligned with the hierarchical signal structure and establish such stronger guarantees based on the HiRIP.  
  Hierarchical measurement operators model multidimensional mixtures of sub-level operators that are expected to arise in many applications. 
  We have seen that if both the mixture procedure as well as all sub-level operators possess the RIP, the hierarchical measurement has the HiRIP. If additionally the sub-level operators are incoherent, the RIP-requirements on the mixture operation can be weakened.  
  These results open up for structured recovery for a class of signals and measurement operators that cannot be captured  to the same extent by classical compressed sensing methods.
  We have demonstrated that the framework can be applied in several mobile communication applications in numerical simulations.  
  We hope to explore further potential applications, and measurement operator classes, in future research.
  
  Finally, a natural generalization of hierarchical sparse signals are signals that combine low-rank and sparsity assumptions on nested hierarchy levels. 
  An simple instance where such structures arise is the de-mixing problem of a sparse sum of low-rank matrices.   
  Generalization of the hierarchical sparse recovery framework to such structures has been recently studied in the context of blind quantum tomography \cite{RothEtAl:2020:Semidevicedependent}.  
  We expect that many of the results presented here can be directly generalized and translated to this setting.

  \paragraph{Acknowledgement} 
  This work has partially been funded by the DFG SPP 1798 - Compressed Sensing in Information Processing. A.~F. acknowledges support from Chalmers AI Research Center (CHAIR).  I.~R. and J.~E. acknowledge funding from the DFG (EI 519/9-1 CoSIP). 
  

  \bibliographystyle{elsarticle-harv}

\bibliography{ACHArefs}
  
  \section{Proofs} \label{sec:proofs}
  
In this section, we present the proofs omitted in the main part of the text.

\subsection{Proof of  Proposition \ref{prop:optimality}}
We begin with the proposition stating that when the sub-level matrices are not incoherent, the top-level matrix needs to have the RIP for the hierarchical measurement operator to have the HiRIP.
\begin{proof}[Proof of Proposition \ref{prop:optimality}]

 Let $S$ with $\abs{S}=s$ be arbitrary. Define, for $\vec c$ arbitrary with $\supp (\vec{c}) \sse S$, the vector $\vec{x}=(\vec{x}_1, \dots, \vec{x}_N)$ with non-vanishing blocks $\vec{x}_i= c_i\vec{g}_i$ for  $i \in S$ and
$\vec{g}_i$ defined in the statement of the proposition.   \af{ Letting $\vec{w}$ be the common value for $\vec{B}_i\vec{g}_i$ for all $i\in S$, the definition of the $\sigma_i$-RIP constants, then implies
\begin{align*}
  \sum_{i\in S} \abs{c_i}^2(1-\delta_{\sigma_i}(\vec{B}_i))\norm{\vec{g}_i}^2 &\leq  \sum_{i\in S} \abs{c_i}^2 \norm{\vec{B}_i\vec{g}_i}^2 \leq \sum_{i\in S} \abs{c_i}^2( 1+\delta_{\sigma_i}(\vec{B}_i)) \norm{\vec{g}_i}^2 
\end{align*}
and, thus, 
\begin{align}
( 1-\sup_i \delta_{\sigma_i}(\vec{B}_i)) \norm{\vec{x}}^2 &\leq \norm{\vec{c}}^2 \norm{\vec{w}}^2  \leq (  1+\sup_i \delta_{\sigma_i}(\vec{B}_i)) \norm{\vec{x}}^2\, . \label{eq:c_x_w}
\end{align}
Now notice that
\begin{align*}
    \vec{H}(\vec{x}) = \sum_{i \in S} \vec{a}_i \otimes (c_i \vec{B}_i \vec{g}_i) = \left(\sum_{i \in S} c_i \vec{a}_i \right) \otimes \vec{w} = \vec{A}\vec{c} \otimes \vec{w}.
\end{align*}
Consequently
\begin{align*}
    \abs{\norm{\vec{A}\vec{c}}^2\norm{\vec{w}}^2 - \norm{\vec{c}}^2\norm{\vec{w}}^2} \leq \abs{ \norm{\vec{H}(\vec{x})}^2- \norm{\vec{x}}^2} + \abs{ \norm{\vec{x}}^2 - \norm{\vec{c}}^2 \norm{\vec{w}}^2} \leq (\delta_{(s,\sigmas)}(\vec{H})  + \sup_{i} \delta_{\sigma_i}(\vec{B}_i)) )\norm{\vec{x}}^2 .
\end{align*}
Dividing the above by $\norm{\vec{w}}^2$, and utilizing \eqref{eq:c_x_w}, we obtain
\begin{align*}
    \abs{\norm{\vec{Ac}}^2-\norm{\vec{c}}^2} \leq (\delta_{(s,\sigmas)}(\vec{H})  + \sup_{i} \delta_{\sigma_i}(\vec{B}_i)) \cdot \frac{\norm{\vec{x}}^2}{\norm{\vec{w}}^2}  \leq \frac{\delta_{(s,\sigmas)}(\vec{H})  + \sup_{i} \delta_{\sigma_i}(\vec{B}_i)}{1-\sup_{i} \delta_{\sigma_i}(\vec{B}_i)} \norm{\vec{c}}^2.
\end{align*}
Since $\vec{c}$ and $S$ were arbitrary, the claim follows.}

\end{proof}

\subsection{Proof of Proposition \ref{prop:incoherentCollections}}

Here, we prove that by drawing matrices $\vec{B}_i$ from ``off-the-shelf'' compressed-sensing ensembles, we obtain pairwise $(\delta,\sigma)$-incoherent collections. Let us first begin with a lemma, which will later let us use standard results on RIP-properties of random matrices.

\begin{lem} \label{lem:offdiag}
     Assume that the family $(\vec{B}_i)_{i \in [N]}$ has the property that for each $i\neq j$, the matrix 
     \begin{align*}
         \vec{C}_{i,j} = [\vec{B}_i, \vec{B}_j] \in \K^{m, n_i + n_j}
     \end{align*}
     has $2\sigma$-RIP constant $\delta_{2\sigma}$ smaller than $\delta^*$. Then, the family is pairwise $(\delta^*, \sigma)$-incoherent.
\end{lem}
\begin{proof}
    Let $\vec{v}_i$ and $\vec{v}_j$ be $\sigma$-sparse and normalized. Let further $\abs{\theta}=1$ be arbitrary. The vector $$\vec{h}=\begin{bmatrix} \vec{v}_i \\  \theta \vec{v}_j\end{bmatrix}\in \K^{2m}$$ is $2\sigma$-sparse. Hence
    \begin{align*}
        \abs{\norm{[ \vec{B}_i, \vec{B}_j ]\vec{h}}^2 - \norm{\vec{h}}^2} \leq \delta^* \norm{\vec{h}}^2.  
    \end{align*}
    It is not hard to see that $\delta_{\sigma}(\vec{B}_i), \delta_\sigma(\vec{B}_j) \leq \af{\delta_{2\sigma}(\vec{C}_{i,j})} \leq \delta^*$. Consequently,
    \begin{align*}
       \norm{[ \vec{B}_i, \vec{B}_j ]\vec{h}}^2 &=  \norm{\vec{B}_i \vec{g}_i}^2 + \norm{\vec{B}_j \vec{g}_j}^2 + 2 \mathrm{Re}(\sprod{\vec{B}_i\vec{g}_i, \theta \vec{B}_j \vec{g}_j}) \geq (1- \delta^*) \norm{\vec{h}}^2 + 2 \mathrm{Re}(\sprod{\vec{B}_i\vec{g}_i, \theta \vec{B}_j \vec{g}_j}).
    \end{align*}
     Now we combine these inequalities and choose $\theta$ such that $\mathrm{Re}(\sprod{\vec{B}_i\vec{g}_i, \theta \vec{B}_j \vec{g}_j}) =  \abs{\sprod{\vec{B}_i\vec{g}_i, \vec{B}_j \vec{g}_j}}$, to obtain
    \begin{align*}
        \abs{\sprod{\vec{B}_i\vec{g}_i,  \vec{B}_j \vec{g}_j}} \leq \delta^*\af{\norm{\vec{h}}^2 = 2\delta^*} \, .
    \end{align*}
\end{proof}

We may now prove the proposition.
\begin{proof}[Proof of Proposition \ref{prop:incoherentCollections}]
    We prove that the two types of random matrices obey the assumption of Lemma \ref{lem:offdiag} with high probability. This will give the claim.
    We remind of the definition $\vec{C}_{i,j}=[\vec{B}_i,\vec{B}_j]$ from the previous lemma.

    1. Notice that for each $i \neq j$, $\vec{C}_{i,j}$ is a matrix with i.i.d sub-Gaussian entries. For a subset $S$ of $[2n]$, let us define $E_{i,j}(S)$ as the event
    \begin{align*}
         \sup_{\substack{\mathrm{supp}(\vec{g}) \sse S, \norm{\vec{g}} \leq 1}}\abs{\norm{\vec{C}_{i,j}\vec{g}}^2 -  \norm{\vec{g}}^2}\geq \delta
    \end{align*}Following a standard proof for the restricted isometry property for sub-Gaussian matrices (e.g., \citep[p.~276-278]{FouRau13})  one proves that for each $S \sse [n]$ with $\abs{S} \leq 2\sigma$
    \begin{align*}
        \mathbb{P}\left(E_{i,j}(S)\right) \leq C_1^{2\sigma} e^{-C_2 \delta^2 m},
    \end{align*}
    where $C_1$ and $C_2$ are constants that only depend on the sub-Gaussian parameters of $\mathscr{D}$. 
    A union bound over multiple events yields
    \begin{align*}
        \mathbb{P}\left(\exists i \neq j, S :  \abs{S}=2\sigma, \ E_{i,j}(S)  \right) \leq N^2\binom{2n}{2\sigma}C_1^\sigma e^{-C_2 \delta ^2 m}.
    \end{align*}
    The last expression is smaller than $\epsilon$ provided
    \begin{align*}
        m \geqsim \delta^{-2} \left( \sigma\log\left(\frac{n}{\sigma}\right)+\log(N) + \log(\epsilon^{-1})\right),
    \end{align*}
    which is the claim. \newline
    
    2. Consider the following $2n$ functions on $\Omega=[n]^2\times\set{-1,1}^2$
    \begin{align*}
        \phi_{\ell}(s,t,\xi,\eta) &= \sqrt{n}\xi U_{s\ell},\\
         \af{\phi_{n+\ell}(s,t,\xi, \eta)} &= \sqrt{n} \eta U_{t\ell},
    \end{align*}
    Then, $(\phi_{\af{1}}, \dots, \phi_{\af{2n}})$ forms a \emph{bounded orthonormal system} on $\Omega$ with respect to the uniform measure $\nu$ in the following sense: For $\af{ \ell_1, \ell_2 \in [n]}$, we have, \af{ using the shorthand $\omega= (s,t,\xi,\eta)$,}
    \begin{align*}
        \int_{\Omega} \phi_{\ell_1}(\omega)\overline{\phi_{\ell_2}(\omega)} d\nu(\omega) & = \frac{1}{n^2}\sum_{s,t=1}^n n U_{t \ell_1} \overline{U_{t \ell_2 }}\cdot \frac{1}{4}\sum_{\xi,\eta \in \set{-1,1}}\xi^2 = \delta_{\ell_1,\ell_2} \\ 
        \int_{\Omega} \phi_{\ell_1+n}(\omega)\overline{\phi_{\ell_2+n}(\omega)} d\nu(\omega)  &= \frac{1}{n^2}\sum_{s,t=1}^n n U_{s\ell_1} \overline{U_{s\ell_2}} \cdot \frac{1}{4}\sum_{\xi,\eta \in \set{-1,1}}\eta^2 = \delta_{\ell_1,\ell_2} \\
        \int_{\Omega} \phi_{\ell_1}(\omega)\overline{\phi_{n+\ell_2}(\omega)} d\nu(\omega) &= \frac{1}{n^2}\left(\sum_{k_1=1}^n U_{s\ell_1 }\right) \left(\sum_{k_2=1}^n\overline{U_{t\ell_2 }}\right) \cdot \frac{1}{4}\sum_{\xi,\eta \in \set{-1,1}}\xi\eta = 0
    \end{align*}
    and it is clear that $\sup_{\ell \in [n], \omega \in \Omega} \abs{\phi_\ell(\omega)}\leq K$. Now we notice that the matrix with $k$:th row equal to
    \begin{align*}
        [\phi_1(\omega_k), \dots \phi_{2n}(\omega_k)]
    \end{align*}
    with $\omega_k$ uniformly sampled on $\Omega$, is distributed exactly as $\vec{C}_{i,j}$. 
    This allows us to use the standard theory of bounded orthonormal systems. Following Ref.\  \citep[p.~404-p.416]{FouRau13}, we see that if we define the event
    \begin{align*}
        F_{i,j}=\sup_{\substack{\norm{\vec{g}}\leq 1 \\
        \vec{g} \ 2\sigma\text{-sparse}}} \abs{\norm{\vec{C}_{i,j}\vec{g}}^2 -  \norm{\vec{g}}^2}\geq \eta_1^2 +\eta_1 +\delta,
    \end{align*}
     where $\eta$ is a parameter which must obey
     \begin{align*}
         \eta_1 \geqsim K \frac{\sqrt{4\sigma}\log(8\sigma)\sqrt{\log(9m)\log(16n)}}{\sqrt{m}}. 
     \end{align*}
     We have (see p.\ 416 of Ref.\ \citep{FouRau13})
    \begin{align*}
        \prb{F_{i,j}} \leq e^{-c_1 \frac{m\delta^2}{K^2\sigma}},
    \end{align*}
    where $c_1$ is a universal constant. A union bound hence implies that as long as
    \begin{align*}
         m &\geqsim K^2 \sigma \delta^{-2} \log(N)\log(\epsilon^{-1}) ,\\
        \frac{m}{\log(9m)} &\geqsim K^2 \sigma \delta^{-2} \log(8\sigma)^2\log(16n),
    \end{align*}
    the collection is $(\delta,\sigma)$-incoherent with a probability bigger than $1-\epsilon$. Putting $\epsilon=n^{-\log(n)^3}$ (and bounding ($\log(m)\log(\sigma)^2 \leqsim \log(n)^3$) yields the claim.
\end{proof}

\subsection{Proof of Theorem \ref{th:Incoherent Blocks}}

We move on to the proof of the second main result of this work, Theorem~\ref{th:Incoherent Blocks}. Let us begin by proving a version of Lemma \ref{lem:SchattenConc} more suitable for the purposes of this section.
\begin{lem} \label{lem:SchattenConc2} Let $\vec{A}\in \K^{M,N}$ have the $2s$-RIP and $X$ be \emph{disjointly $s$-bisparse} matrix. That is, assume that there exists two disjoint subsets of $S$, $S'$ of $[N]$ with $\abs{S}, \abs{S'}\leq s$ so that
\begin{align*}
    X_{i,j}=0 \text{ if  $i \notin S$ or $j \notin S'$}\, .
\end{align*}
    Then
    \begin{align*}
        \abs{\sprod{\vec{A}^*\vec{A},\vec{X}}} \leq \delta_{2s}(\vec{A})\norm{\vec{X}}_*\, .
    \end{align*}
\end{lem}
\begin{proof}
We use exactly the same idea as in the proof of Lemma \ref{lem:SchattenConc}. We may form a singular value decomposition of $\vec{X}$
\begin{align*}
    \sum_{i=1}^N \rho_i \vec{x}_i \vec{y}_i^*\, ,
\end{align*}
where $\vec{x}_i$ and $\vec{y}_i$ are left and right normalized singular vectors of $\vec{x}_i$ and $\vec{y}_i$, and $\rho_i$ are singular values of $\vec{X}$. Now, since $\vec{X}$ is disjointly $s$-bisparse, there must be $\supp{\vec{x}_i} \sse S$ and $\supp{\vec{y}_i} \sse S'$ whenever $\rho_i \neq 0$. We get
\begin{align*}
   \abs{ \sprod{\vec{A}^*\vec{A},\vec{X}} }\leq \sum_{i} \rho_i  \abs{\sprod{\vec{A}_i \vec{x}_i, \vec{A} \vec{y}_i}}.
\end{align*}
It is a well known result that the disjointness of $S$ and $S'$ (see, e.g., Ref.\ \citep[Prop. 6.3]{FouRau13}) implies that 
\begin{align*}
    \abs{\sprod{\vec{A} \vec{x}_i, \vec{A} \vec{y}_i}} \leq \delta_{2s}(\vec{A}).
\end{align*}
The claim follows.

\end{proof}

Now, we have all the tools in place to prove the theorem.
\begin{proof}[Proof of Theorem \ref{th:Incoherent Blocks}]
We start just as in the proof of Theorem~\ref{th:HiRIP}: Let $\vec x=(\vec x_1, \ldots, \vec x_N)$ be $(\lambda s, \sigmas)$-sparse and normalized. We then have
\begin{align*}
    \norm{\vec{H}(\vec x)}^2 = \sprod{\vec{A}^*\vec{A}, \vec{G}}
\end{align*}
where $G_{i,j}= \sprod{\vec{B}_i\vec{x}_i, \vec{B}_j \vec{x}_j}$. Now let us divide the block support of $\vec{x}$ into $\lambda$ disjoint groups $S_k$, each of cardinality $s$. Then define $\vec{G}^{k,\ell}$ through
\begin{align*}
    G^{k,\ell}_{i,j}= \begin{cases} \sprod{\vec{B}_i\vec{x}_i, \vec{B}_j \vec{x}_j} & \text{ if $i \in S_k$ and $j\in S_\ell$ } \\
    0 & \text{ else}.
    \end{cases}
\end{align*}
    Then $\vec{G}^{k,\ell}$ is $s$-bisparse for $k=\ell$ and disjointly $s$-bisparse  for $k \neq \ell$. Thus, applying Lemmata \ref{lem:SchattenConc} and \ref{lem:SchattenConc2} yields for $k \neq \ell$
    \begin{align*}
        \abs{\sprod{\vec{A}^*\vec{A}, \vec{G}^{k,k}} - \norm{\vec{G}^{k,k}}_*} &\leq \delta_s(\vec{A}) \norm{\vec{G}^{k,k}}_*, \\
        \abs{\sprod{\vec{A}^*\vec{A}, \vec{G}^{k,\ell}}} &\leq \delta_{2s}(\vec{A}) \norm{\vec{G}^{k,\ell}}_*\, .
    \end{align*}
    Consequently,
    \begin{align*}
        \abs{ \sprod{\vec{A}^*\vec{A}, \vec{G}} - \sum_{k} \norm{\vec{G}^{k,k}}_*} \leq &  \sum_{k} \delta_s(\vec{A}) \norm{\vec{G}^{k,k}}_* 
          + \sum_{k \neq \ell} \delta_{2s}(\vec{A}) \norm{\vec{G}^{k,\ell}}_*\, .
    \end{align*}
    We deal with the diagonal terms just as in the proof of Theorem~\ref{th:HiRIP}: We obtain
    \begin{align*}
        \abs{\norm{\vec{G}^{k,k}}_* - \sum_{i \in S_k} \norm{\vec{x}_i}^2} \leq \delta_{\sigma}^*\sum_{i \in S_k} \norm{\vec{x}_i}^2.
    \end{align*}
    By summing over $k$, we have
    \begin{align*}
        \abs{ \sprod{\vec{A}^*\vec{A}, \vec{G}} - 1} \leq& \delta_s(\vec{A}) + \delta_\sigma^* + \delta_s(\vec{A})\delta_{\sigma}^* 
        +\sum_{k \neq \ell} \delta_{2s}(\vec{A}) \norm{\vec{G}^{k,\ell}}_*\, .
    \end{align*}
    Now we need to analyse the cross diagonal terms. Since each term is a matrix of rank at most $s$, we have
    \begin{align*}
        \norm{\vec{G}^{k,\ell}}_* \leq \sqrt{s}\norm{\vec{G}^{k,\ell}}_F.
    \end{align*}
    The $(\delta,\sigma)$-incoherence assumption gives us the entry-wise bound $\abs{G_{i,j}^{k,\ell}}^2 \leq \delta_{2\sigma^*} \norm{\vec{x}_i}^2 \norm{\vec{x}_j}^2$ on $\vec{G}^{k,\ell}$. That implies
    \begin{align*}
        \norm{\vec{G}^{k,\ell}}_F^2  \leq \sum_{i \in S_k} \sum_{j\in S_\ell} \delta_{2\sigma^*}^2 \norm{\vec{x}_i}^2 \norm{\vec{x}_j}^2.
    \end{align*}
    Therefore, 
    \begin{align*}
        \sum_{k \neq \ell} \delta_{2s}(\vec{A}) \norm{\vec{G}^{k,\ell}}_* &\leq \sqrt{s} \delta_{2\sigma}^* \sum_{k \neq \ell} \left(\sum_{i \in S_k}\norm{\vec{x}_i}^2 \right)^{\tfrac12}  \left(\sum_{j\in S_\ell} \norm{\vec{x}_j}^2\right)^{\tfrac12} \\
        & \leq \sqrt{s} \delta_{2\sigma}^* \left(\sum_{k}\left(\sum_{i \in S_k}\norm{\vec{x}_i}^2 \right)^{\tfrac12}\right)^2 \leq \lambda \sqrt{s} \delta_{2\sigma}^*\, ,
    \end{align*}
    where we used Cauchy-Schwarz and the (implicit) normalization assumption in the final line.

\end{proof}
\subsection{Proof of Proposition~\ref{prop:s_vs_delta}}

Here, we give the proof of  Proposition~\ref{prop:s_vs_delta}, the special case of the  `deciphering' Theorem~\ref{th:Incoherent Blocks} for a Gaussian top-level matrix.
\begin{proof}[Proof of Proposition~\ref{prop:s_vs_delta}] It is well known that (see, e.g., Ref.\ \citep[Proof of Th. 9.11]{FouRau13})
\begin{align*}
    \prb{\text{$\vec{A}$ does not have the $(t,\delta)$-RIP}} \leq C_1^{t} \binom{N}{t} e^{-C_2 M\delta^2}.
\end{align*}
\af{Now note that if $\vec{A}$ has the $\left(s, \tfrac{\delta}{2(1+\delta_\sigma^*)}\right)$-RIP and $\left(2s,\tfrac{\delta \sqrt{s}}{2t\delta_{2\sigma}^*}\right)$-RIP, Theorem \ref{th:Incoherent Blocks}, with $\lambda = \sfrac{t}{s}$, implies that
\begin{align*}
    \delta_{(t,\sigmas)}(\vec{H}) \leq \delta_s(\vec{A})(1+\delta_\sigma^*) + \delta_{\sigma}^* + \tfrac{t}{s} \cdot \sqrt{s}\delta_{2s}(\vec{A})\delta_{2\sigma}^* \leq \tfrac{\delta}{2} + \delta_{\sigma}^* + \tfrac{\delta}{2} = \delta_{\sigma}^* + \delta.
\end{align*}
Consequently,
\begin{align*}
    &\prb{\text{$\vec{H}$ does not have the $\left((t,\sigmas),(\delta+\delta_{\sigma}^*)\right)$-HiRIP}} \\ & \qquad\leq   \prb{ \text{$\vec{A}$ does not have the $\left(s,\frac{\delta}{2(1+\delta_\sigma^*)}\right)$-RIP}} 
    + \prb{ \text{$\vec{A}$ does not have the $\left(2s,\frac{\sqrt{s}\delta}{2t\delta_{2\sigma}^*}\right)$-RIP}} \\
    & \qquad \qquad\leq 
    C_1^{s} \binom{N}{s} e^{-\frac{C_2 M\delta^2}{4(1+\delta_\sigma^*)^2}} + C_1^{2s} \binom{N}{2s} e^{-C_2 \frac{Mt^2(\delta_{2\sigma}^*)^2\delta^2}{4s}}.
\end{align*}
It suffices to make each term smaller than $\epsilon/2$ for any chosen $\epsilon>0$. Sufficient for this is
\begin{align*}
    M \geq \frac{4(1+\delta^*_\sigma)^{2}}{C_2\delta^2}\left(s \log(C_1) + \log\binom{N}{s}+\log(2\epsilon^{-1})\right), \\
    M \geq \frac{4s}{C_2 t^2(\delta_{2\sigma}^*)^2\delta^{2}}\left(2s \log(C_1) + \log\binom{N}{2s} + \log(2\epsilon^{-1})\right).
\end{align*}
Using the asymptotic estimate $\log\binom{N}{s} \sim s \log\frac{N}{s}$, we obtain the following sufficient conditions
\begin{align*}
 M \geqsim \frac{4(1+\delta^*_\sigma)^{2}s}{C_2\delta^2} \cdot \left( \log(C_1) + \log\left(\frac{N}{s}\right) + s^{-1}\log(2\epsilon^{-1})\right), \\
     M \geqsim \frac{4s^2}{C_2 t^2(\delta_{2\sigma}^*)^2\delta^{2}} \left( \log(C_1) + \log\left(\frac{N}{2s}\right)+ s^{-1}\log(2\epsilon^{-1})\right) .
\end{align*}
If we now choose $s \sim (t\delta_{2\sigma}^*)^2$, both equations above turn in to
\begin{align*}
    M \geqsim \frac{(t\delta_{2\sigma}^*)^2}{C_2\delta^{2}} \left( \log(C_1) + \log\left(\frac{N}{2s}\right)+ s^{-1}\log(2\epsilon^{-1})\right) ,
\end{align*}
which is the assumption of the proposition. The proof is finished.}
\end{proof}

\end{document}